\documentclass{article}
\usepackage{eurosym}
\usepackage{amssymb}
\usepackage{amsmath}
\usepackage{cite}
\usepackage{graphicx}
\usepackage{epsf}
\usepackage{lscape}
\usepackage{epstopdf}

\input{tcilatex}

\begin{document}

\title{Shallow-water equations with complete Coriolis force: Group
Properties and Similarity Solutions}
\author{Andronikos Paliathanasis\thanks{%
Email: anpaliat@phys.uoa.gr} \\
{\ \textit{Institute of Systems Science, Durban University of Technology }}\\
{\ \textit{PO Box 1334, Durban 4000, Republic of South Africa}}}
\maketitle

\begin{abstract}
The group properties of the shallow-water equations with the complete
Coriolis force is the subject of this study. In particular we apply the Lie
theory to classify the system of three nonlinear partial differential
equations according to the admitted Lie point symmetries. For each case of
the classification problem the one-dimensional optimal system is determined.
The results are applied for the derivation of new similarity solutions. 
\newline
\newline
\newline

\end{abstract}

\section{Introduction}

\label{sec1}

Shallow-water equations are a set of hyperbolic nonlinear partial
differential equations, known also as the Saint-Venant equations, which
describe a thin layer of inviscid fluid with a free surface. There are
various applications of the Shallow-water equations in physical science, for
some recent results we refer the reader to \cite%
{sww01,sww02a,sww02,sww03,sww05,sww06} and references therein. In general
hyperbolic equations are of special interest for the study of flows.
Two-phase flow is of special interest because it is described by a set of
hyperbolic equations with non-equilibrium phenomena between different
phases. There are various real world applications for these models, hence
there are various studies in the literature \cite%
{tw1,tw2,tw3,tw4,tw5,tw6,tw7,tw8}.

In this piece of work we are interested in the group properties of the
rotating two-dimensional shallow-water equations. The shallow-water
equations with the complete Coriolis force term and topography were derived
in \cite{cor1}.

In the case of a linear plane the shallow-water equations are

\begin{equation}
h_{t}+\left( hu\right) _{x}+\left( hv\right) _{y}=0,  \label{sw.01}
\end{equation}%
\begin{equation}
u_{t}+uu_{x}+vu_{x}-\left( 2\Omega _{z}-\Omega _{y}h_{y}\right) v+\left(
gh-h^{2}\Omega _{y}u\right) _{x}-\Omega _{y}\left( \left( hu\right)
_{x}+\left( hv\right) _{y}\right) =0,  \label{sw.02}
\end{equation}%
\begin{equation}
v_{t}+uv_{x}+vv_{y}+\left( 2\Omega _{z}-\Omega _{y}h_{y}\right) u+\left(
gh-h^{2}\Omega _{y}u\right) _{y}=0,  \label{sw.03}
\end{equation}%
where the axes have been selected to be tangent to a sphere\ at a specific
latitude $\phi _{0}~$such that the Coriolis term $\Omega =\left( \Omega
_{x},\Omega _{y},\Omega _{z}\right) $ has components $\Omega _{x}=0,~\Omega
_{y}=\Omega \cos \phi _{0}~,~\Omega _{z}=\Omega \sin \phi _{0}$.

We study the one-parameter point transformations which leave invariant the
set of the three equations, (\ref{sw.01}), (\ref{sw.02}) and (\ref{sw.03}).
Specifically, the theory of Lie point symmetries is applied in order to
determine the admitted Lie point symmetries and the one-dimensional optimal
system.

Lie theory is a powerful mathematical treatment for the study of nonlinear
differential equations. The existence of a transformation which leaves
invariant a set of differential equations indicates the existence of
invariant functions, known also as Lie invariants, which can be used to
simplify the set of differential equations by reduction of the differential
equations. There are various applications of the Lie point symmetries in
physical theories \cite{ao01,ao02,ao03,ao04,ao05,ao06,ao07,ao08} as also in
fluid dynamics and in shallow-water equations. The Lie point symmetries of
the non-rotating shallow-water equations were studied in \cite{sw3}, while a
similar analysis with specific Coriolis term was performed in \cite{swr06}.
In the latter case it was found that the special Coriolis term can be
neglected by the field equations with the use of point transformations.
Rotating shallow-water equations with more general fluids were studied in 
\cite{ans1} while other studies of shallow-water equations can be found in 
\cite{sw1,sw2,sw3,sw4,sw5,sw6,sw7,sw9,sw10} and in some recent works in \cite%
{mel1,mel2,mel3,mel4,mel5}. In addition, for the hyperbolic equations which
describe the two-phase flow Lie symmetries were applied in \cite{tw9,tw10}.

Ovsiannikov \cite{Ovsi} in 1982 demonstrated the construction of the
one-dimensional optimal system for the Lie algebra, using a global matrix
for the adjoint transformation. Since then the classification of the
one-dimensional optimal system has become a main tool for the study of
nonlinear differential equations \cite{opt1,opt2,opt3}. Such analysis for
the shallow-water system has been presented before in \cite%
{swr06,ans1,ss01,ss02}. However, in this study we present for the first time
the complete symmetry classification for the shallow-water equations with a
complete Coriolis term. The structure of the paper is as follows.

In Section \ref{sec2} we present the basic properties and definitions of
Lie's theory. The classification of the Lie point symmetries for the system
of our study is presented in Section \ref{sec3}. We find that in the extreme
values of the latitude $\phi _{0}$ the shallow-water equations admit
additional Lie point symmetries, while when $\phi _{0}=\frac{\pi }{2}$ we
recover previous results of the literature. For all the different cases, we
classify the one-dimensional optimal system which is essential to perform
reductions and calculate similarity solutions. In Section \ref{sec4} we
apply the Lie invariants to reduce the dependent variables of the
shallow-water equations and we present the cases for which similarity
solutions can be expressed in terms of closed-form function. Finally in
Section \ref{sec5} we summarize our results and we draw our conclusions.

\section{Preliminaries}

\label{sec2}

We present the basic definitions and properties for the mathematical tools
which we apply in this work. Precisely we give the main definition of point
symmetries and of a Lie symmetry vector. Consider now the system $H^{A} $ of
partial differential equations with independent variables $y^{i}$ and
dependent variables $u^{A}=u^{A}\left( y^{i}\right) ,$~that is%
\begin{equation}
H^{A}\left( y^{i},u^{A},u_{i}^{A},...\right) \equiv 0,
\end{equation}%
where $u_{i}^{A}=\frac{\partial u^{A}}{\partial y^{i}}$.

We define the infinitesimal one-parameter point transformation 
\begin{eqnarray}
\bar{y}^{i} &=&y^{i}\left( y^{j},u^{B};\varepsilon \right) ,  \label{de.02}
\\
\bar{u}^{A} &=&u^{A}\left( y^{j},u^{B};\varepsilon \right) ,  \label{de.03}
\end{eqnarray}%
defined in the space of variables $\left\{ y^{i},u^{A}\right\} ~$in $%
\varepsilon $ is an infinitesimal parameter. From (\ref{de.02}), (\ref{de.03}%
) the infinitesimal transformation is defined%
\begin{eqnarray}
\bar{y}^{i} &=&y+\varepsilon \xi ^{i}\left( y^{j},u^{B}\right) \\
\bar{u}^{A} &=&u^{A}+\varepsilon \eta ^{A}\left( y^{j},u^{B}\right)
\end{eqnarray}%
with extension in the jet space $\left\{ y^{i},u^{A},u_{,i}^{A},...\right\} $
as%
\begin{eqnarray*}
\bar{u}_{i}^{A} &=&u_{i}^{A}+\varepsilon \eta _{i}^{A\left[ 1\right] } \\
&&... \\
\bar{u}_{i_{1}i_{2}...i_{n}}^{A} &=&u_{i_{1}i_{2}...i_{n}}^{A}+\varepsilon
D_{i}\eta ^{A}-u_{i_{1}i_{2}...i_{n-1}}^{A}D_{i}\left( \xi ^{j}\right),
\end{eqnarray*}%
where%
\begin{equation}
\eta _{i}^{A\left[ 1\right] }=D_{i}\eta
^{A}-u_{i_{1}i_{2}...i_{n-1}}D_{i}\left( \xi ^{j}\right)
\end{equation}%
and%
\begin{equation}
\eta ^{A\left[ n\right] }=D_{i}\eta ^{A\left[ n-1\right]
}-u_{i_{1}i_{2}...i_{n-1}}D_{i}\left( \xi ^{j}\right) .
\end{equation}

Under the action of the one-parameter point transformation (\ref{de.02}), (%
\ref{de.03}) the system of differential equations becomes $\bar{H}^{A}\left( 
\bar{y}^{i},\bar{u}^{A},\bar{u}_{i}^{A},...,\varepsilon \right) $. We say
that the system $H^{A}$ remains invariant under the action of the
one-parameter point transformation (\ref{de.02}), (\ref{de.03}) if and only
if $\bar{H}^{A}\left( \bar{y}^{i},\bar{u}^{A},\bar{u}_{i}^{A},...,%
\varepsilon \right) $ $\equiv 0$, or equivalently \cite{olver,kumei,ibra}%
\begin{equation}
\lim_{\varepsilon \rightarrow 0}\frac{\bar{H}^{A}\left( \bar{y}^{i},\bar{u}%
^{A},...;\varepsilon \right) -H^{A}\left( y^{i},u^{A},...\right) }{%
\varepsilon }=0,  \label{ls.05}
\end{equation}
that is, 
\begin{equation}
\mathcal{L}_{X}\left( H^{A}\right) =0,  \label{ls.05a}
\end{equation}%
where~$\mathcal{L}$ denotes the Lie derivative with respect the generator $%
X~ $of the infinitesimal transformation (\ref{de.02}), (\ref{de.03}), that
is,%
\begin{equation*}
X=\xi ^{i}\left( y^{j},u^{B}\right) \partial _{i}+\eta ^{A}\left(
y^{j},u^{B}\right) \partial _{A}.
\end{equation*}%
The latter vector field is set in the jet space, $\left\{
y^{i},u^{A},u_{,i}^{A},...\right\}, $ 
\begin{equation}
X^{\left[ n\right] }=X+\eta ^{\left[ 1\right] }\partial
_{u_{i}^{A}}+...+\eta ^{\left[ n\right] }\partial
_{u_{i_{i}i_{j}...i_{n}}^{A}}.  \label{ls.06}
\end{equation}

When (\ref{ls.05a}) is true for a specific vector field $X$, then the vector
field $X$ is a Lie symmetry for the system of partial differential equations 
$H^{A}$.

The admitted symmetry vectors of a given set of differential equations
constitute a closed-group known as a Lie group. A Lie group defines the main
properties for the set of the differential equations. The main application
of the Lie symmetries is the determination of solutions known as similarity
solutions and follow from the application of the Lie invariants in the
differential equations. However, in order to classify all the possible
similarity transformations and solutions the one-dimensional optimal system
should be calculated \cite{olver}.

\section{Lie symmetry analysis}

\label{sec3}

We apply the theory for the system of the three-dimensional partial
differential equations, (\ref{sw.01}), (\ref{sw.02}) and (\ref{sw.03}). It
follows that the admitted Lie point symmetries for arbitrary latitude $\phi
_{0}$ are 
\begin{equation*}
X_{1}=\partial _{t}~,~X_{2}=\partial _{x}~,~X_{3}=\partial _{y},
\end{equation*}%
In Table \ref{tab1} we present the commutators for the three-dimensional Lie
algebra, $\left\{ X_{1},X_{2},X_{3}\right\} $. The adjoint representation of
the latter Lie algebra is presented in Table \ref{tab2}. From Table \ref%
{tab1} we infer that the admitted Lie algebra is the $3A_{1}$ according to
the classification scheme of Patera et al. \cite{pat11,pat12}

From tables \ref{tab1} and \ref{tab2} we conclude that the one-dimensional
optimal system \cite{olver} of the four-dimensional Lie algebra consists of
the vector fields 
\begin{equation*}
\left\{ X_{1}\right\} ~,~\left\{ X_{2}\right\} ~,~\left\{ X_{3}\right\}
~,~\left\{ X_{1}+\alpha X_{2}\right\} ~,
\end{equation*}%
\begin{equation*}
\left\{ X_{1}+\alpha X_{3}\right\} ~,\left\{ X_{2}+\alpha X_{3}\right\}
,\left\{ X_{1}+\alpha X_{2}+\beta X_{3}\right\} .
\end{equation*}

\begin{table}[tbp] \centering%
\caption{Commutator table for the Lie point symmetries of the Shallow-Water
system}%
\begin{tabular}{cccc}
\hline\hline
$\left[ X_{I},X_{J}\right] $ & $\mathbf{X}_{1}$ & $\mathbf{X}_{2}$ & $%
\mathbf{X}_{3}$ \\ \hline
$\mathbf{X}_{1}$ & $0$ & $0$ & $0$ \\ 
$\mathbf{X}_{2}$ & $0$ & $0$ & $0$ \\ 
$\mathbf{X}_{3}$ & $0$ & $0$ & $0$ \\ \hline\hline
\end{tabular}%
\label{tab1}%
\end{table}%

\begin{table}[tbp] \centering%
\caption{Adjoint representation for the Lie point symmetries of the
Shallow-Water System}%
\begin{tabular}{cccc}
\hline\hline
$Ad\left( e^{\left( \varepsilon \mathbf{X}_{i}\right) }\right) \mathbf{X}%
_{j} $ & $\mathbf{X}_{1}$ & $\mathbf{X}_{2}$ & $\mathbf{X}_{3}$ \\ \hline
$\mathbf{X}_{1}$ & $X_{1}$ & $X_{2}$ & $X_{3}$ \\ 
$\mathbf{X}_{2}$ & $X_{1}$ & $X_{2}$ & $X_{3}$ \\ 
$\mathbf{X}_{3}$ & $X_{1}$ & $X_{2}$ & $X_{3}$ \\ \hline\hline
\end{tabular}%
\label{tab2}%
\end{table}%

For special values of the parameters $\Omega _{z}$ or $\Omega _{y}$ the
resulting Shallow-water system is invariant under a different Lie algebra.
We find that the two limits are for the latitudes (i) $\phi _{0}=0$ and (ii) 
$\phi _{0}=\frac{\pi }{2}$, which correspond to values (i) $\Omega
_{y}=\Omega ~,~\Omega _{z}=0~$, and (ii) $\Omega _{y}=0,~\Omega _{z}=\Omega $%
.

\subsection{Lie symmetries at the equator}

For latitude $\phi _{0}=0$, that is, for the equator, the resulting
Shallow-water equations, (\ref{sw.01}), (\ref{sw.02}) and (\ref{sw.03}), are
written as%
\begin{equation}
h_{t}+\left( hu\right) _{x}+\left( hv\right) _{y}=0,
\end{equation}%
\begin{equation}
u_{t}+uu_{x}+vu_{x}+\left( \Omega _{y}h_{y}\right) v+\left( gh-h^{2}\Omega
_{y}u\right) _{x}-\Omega _{y}\left( \left( hu\right) _{x}+\left( hv\right)
_{y}\right) =0,
\end{equation}%
\begin{equation}
v_{t}+uv_{x}+vv_{y}-\left( \Omega _{y}h_{y}\right) u+\left( gh-h^{2}\Omega
_{y}u\right) _{y}=0
\end{equation}%
and admit the following Lie point symmetries%
\begin{equation*}
Y_{1}=\partial _{t}~,~Y_{2}=\partial _{x}~,~Y_{3}=\partial
_{y},~Y_{4}=t\partial _{t}+x\partial _{x}+y\partial _{y}~,~Y_{5}=t\partial
_{y}+\partial _{V}~
\end{equation*}%
with commutators and adjoint representation as given in Tables \ref{tab3}
and \ref{tab4}. The Lie symmetries form the $\left\{ 3A_{1}\otimes
_{s}2A_{1}\right\} $ Lie algebra.

With the use of the adjoint representation from Table \ref{tab4}~we can
determine the one-dimensional optimal system, that is,%
\begin{equation*}
\left\{ Y_{1}\right\} ~,~\left\{ Y_{2}\right\} ~,~\left\{ Y_{3}\right\}
~,~\left\{ Y_{4}\right\} ~,~\left\{ Y_{5}\right\} ~,~
\end{equation*}%
\begin{equation*}
\left\{ Y_{1}+a_{2}Y_{2}\right\} ~,~\left\{ Y_{1}+a_{3}Y_{3}\right\}
~,~\left\{ Y_{1}+a_{4}Y_{4}\right\} ~,
\end{equation*}%
\begin{equation*}
~~~\left\{ Y_{1}+a_{5}Y_{5}\right\} ~,~\left\{ Y_{2}+a_{3}Y_{3}\right\}
~,~\left\{ Y_{1}+a_{5}Y_{5}\right\} ,
\end{equation*}%
\begin{equation*}
~\left\{ Y_{1}+a_{2}Y_{2}+a_{3}Y_{3}\right\} ~,~\left\{
a_{1}Y_{1}+a_{4}Y_{4}+a_{5}Y_{5}\right\} .
\end{equation*}

\begin{table}[tbp] \centering%
\caption{Commutator table for the Lie point symmetries of the Shallow-Water
	System for $\phi_{0}=0$}%
\begin{tabular}{cccccc}
\hline\hline
$\left[ \mathbf{Y}_{I},\mathbf{Y}_{J}\right] $ & $\mathbf{Y}_{1}$ & $\mathbf{%
Y}_{2}$ & $\mathbf{Y}_{3}$ & $\mathbf{Y}_{4}$ & $\mathbf{Y}_{5}$ \\ \hline
$\mathbf{Y}_{1}$ & $0$ & $0$ & $0$ & $Y_{1}$ & $Y_{3}$ \\ 
$\mathbf{Y}_{2}$ & $0$ & $0$ & $0$ & $Y_{2}$ & $0$ \\ 
$\mathbf{Y}_{3}$ & $0$ & $0$ & $0$ & $Y_{3}$ & $0$ \\ 
$\mathbf{Y}_{4}$ & $-Y_{1}$ & $-Y_{2}$ & $-Y_{3}$ & $0$ & $0$ \\ 
$\mathbf{Y}_{5}$ & $-Y_{3}$ & $0$ & $0$ & $0$ & $0$ \\ \hline\hline
\end{tabular}%
\label{tab3}%
\end{table}%

\begin{table}[tbp] \centering%
\caption{Adjoint representation for the Lie point symmetries of the
Shallow-water System for $\phi_{0}=0$}%
\begin{tabular}{cccccc}
\hline\hline
$Ad\left( e^{\left( \varepsilon \mathbf{Y}_{i}\right) }\right) \mathbf{Y}%
_{j} $ & $\mathbf{Y}_{1}$ & $\mathbf{Y}_{2}$ & $\mathbf{Y}_{3}$ & $\mathbf{Y}%
_{4}$ & $\mathbf{Y}_{5}$ \\ \hline
$\mathbf{Y}_{1}$ & $Y_{1}$ & $Y_{2}$ & $Y_{3}$ & $Y_{4}-\varepsilon X_{1}$ & 
$Y_{5}-\varepsilon X_{3}$ \\ 
$\mathbf{Y}_{2}$ & $Y_{1}$ & $Y_{2}$ & $Y_{3}$ & $Y_{4}-\varepsilon X_{2}$ & 
$Y_{5}$ \\ 
$\mathbf{Y}_{3}$ & $Y_{1}$ & $Y_{2}$ & $Y_{3}$ & $Y_{4}-\varepsilon X_{3}$ & 
$Y_{5}$ \\ 
$\mathbf{Y}_{4}$ & $e^{\varepsilon }Y_{1}$ & $e^{\varepsilon }Y_{2}$ & $%
e^{\varepsilon }Y_{3}$ & $Y_{4}$ & $Y_{5}$ \\ 
$\mathbf{Y}_{5}$ & $Y_{1}+\varepsilon X_{3}$ & $Y_{2}$ & $Y_{3}$ & $Y_{4}$ & 
$Y_{5}$ \\ \hline\hline
\end{tabular}%
\label{tab4}%
\end{table}%

\subsection{Lie symmetries at the pole}

For latitudes near to the pole, i.e. $\phi =\frac{\pi }{2}$, the
Shallow-water equations (\ref{sw.01}), (\ref{sw.02}) and (\ref{sw.03}) can
be written in the simple form 
\begin{equation}
h_{t}+\left( hu\right) _{x}+\left( hv\right) _{y}=0,  \label{s11}
\end{equation}%
\begin{equation}
u_{t}+uu_{x}+vu_{x}+gh_{x}-2\Omega _{z}v=0,  \label{s12}
\end{equation}%
\begin{equation}
v_{t}+uv_{x}+vv_{y}+gh_{y}+2\Omega _{z}u=0.  \label{s13}
\end{equation}%
The group properties of the latter system were studied before in \cite{swr06}%
, where in comparison with \cite{swr06} $f=2\Omega _{z}$.

Equations (\ref{s11}), (\ref{s12}) and (\ref{s13}) are invariant under the
action of a nine-dimensional Lie algebra with elements 
\begin{equation*}
Z_{1}=\partial _{t}~,~Z_{2}=\partial _{x}~,~Z_{3}=\partial _{y}~,
\end{equation*}%
\begin{equation*}
~Z_{4}=x\partial _{x}+y\partial _{y}+u\partial _{u}+v\partial
_{v}+2h\partial _{h}~
\end{equation*}%
\begin{equation*}
Z_{5}=y\partial _{x}-x\partial _{y}+v\partial _{u}-u\partial _{v}~,
\end{equation*}%
\begin{equation*}
Z_{6}=\sin \left( 2\Omega t\right) \partial _{x}+\cos \left( 2\Omega
t\right) \partial _{y}+2\Omega \left( \cos \left( 2\Omega t\right) \partial
_{u}-\sin \left( 2\Omega t\right) \partial _{v}\right) ~,
\end{equation*}%
\begin{equation*}
Z_{7}=-\cos \left( 2\Omega t\right) \partial _{x}+\sin \left( 2\Omega
t\right) \partial _{y}+2\Omega \left( \sin \left( 2\Omega t\right) \partial
_{u}+\cos \left( 2\Omega t\right) \partial _{v}\right) ~,
\end{equation*}%
\begin{eqnarray*}
Z_{8} &=&\sin \left( 2\Omega t\right) \partial _{t}+\Omega \left( x\cos
\left( 2\Omega t\right) +y\sin \left( 2\Omega t\right) \right) \partial
_{x}-\Omega \left( x\sin \left( 2\Omega t\right) -y\cos \left( 2\Omega
t\right) \right) \partial _{y}+ \\
&&-\Omega \left( u\cos \left( 2\Omega t\right) -v\sin \left( 2\Omega
t\right) -2\Omega \left( y\cos \left( 2\Omega t\right) -x\sin \left( 2\Omega
t\right) \right) \right) \partial _{u}+ \\
&&-\Omega \left( u\sin \left( 2\Omega t\right) +u\cos \left( 2\Omega
t\right) +2\Omega \left( y\sin \left( 2\Omega t\right) +x\cos \left( 2\Omega
t\right) \right) \right) \partial _{v}-2\Omega h\cos \left( 2\Omega t\right)
\partial _{h},
\end{eqnarray*}%
\begin{eqnarray*}
Z_{9} &=&\cos \left( 2\Omega t\right) \partial _{t}+\Omega \left( y\cos
\left( 2\Omega t\right) -x\sin \left( 2\Omega t\right) \right) \partial
_{x}-\Omega \left( x\cos \left( 2\Omega t\right) +y\sin \left( 2\Omega
t\right) \right) \partial _{y}+ \\
&&+\Omega \left( u\sin \left( 2\Omega t\right) +v\cos \left( 2\Omega
t\right) -2\Omega \left( x\cos \left( 2\Omega t\right) +y\sin \left( 2\Omega
t\right) \right) \right) \partial _{u}+ \\
&&+\Omega \left( -u\cos \left( 2\Omega t\right) +v\sin \left( 2\Omega
t\right) +\Omega \left( x\sin \left( 2\Omega t\right) -y\cos \left( 2\Omega
t\right) \right) \right) \partial _{v}+2\Omega \sin \left( 2\Omega t\right)
\partial _{h}.
\end{eqnarray*}%
The commutators are presented in Table \ref{tab5}.

The symmetry vectors form a nine-dimensional Lie algebra. Furthermore, the
adjoint representation of the nine-dimensional Lie algebra is given in
Tables \ref{tab6} and \ref{tab6b}.

The shallow-water equations without a Coriolis term admit also nine Lie
point symmetries with the same commutators. Hence, the same Lie algebra with
that of the Coriolis term at the pole but in a different representation.
That common property has been observed before in \cite{swr06}. The
coordinate transformation which relates the two different representations of
the Lie algebra can be use to remove the Coriolis term from the system (\ref%
{s11}), (\ref{s12}) and (\ref{s13}).

We do not continue with the determination of the one-dimensional optimal
system for the system (\ref{s11}), (\ref{s12}) and (\ref{s13}), because it
can be find in \cite{swr06}.

\begin{landscape}
\begin{table}[tbp] \centering%
\caption{Commutator table for the Lie point symmetries of the Shallow-water
system for $\phi_{0}=\pi/2$}%
\begin{tabular}{cccccccccc}
\hline\hline
$\left[ \mathbf{Z}_{I},\mathbf{Z}_{J}\right] $ & $\mathbf{Z}_{1}$ & $\mathbf{%
Z}_{2}$ & $\mathbf{Z}_{3}$ & $\mathbf{Z}_{4}$ & $\mathbf{Z}_{5}$ & $\mathbf{Z%
}_{6}$ & $\mathbf{Z}_{7}$ & $\mathbf{Z}_{8}$ & $\mathbf{Z}_{9}$ \\ \hline
$\mathbf{Z}_{1}$ & $0$ & $0$ & $0$ & $0$ & $0$ & $-2\Omega Z_{7}$ & $2\Omega
Z_{8}$ & $2\Omega Z_{9}$ & $-2\Omega Z_{8}$ \\ 
$\mathbf{Z}_{2}$ & $0$ & $0$ & $0$ & $Z_{2}$ & $-Z_{3}$ & $0$ & $0$ & $%
-2\Omega Z_{7}$ & $-2\Omega Z_{6}$ \\ 
$\mathbf{Z}_{3}$ & $0$ & $0$ & $0$ & $Z_{3}$ & $Z_{2}$ & $0$ & $0$ & $%
\,-2\Omega Z_{6}$ & $-2\Omega Z_{7}$ \\ 
$\mathbf{Z}_{4}$ & $0$ & $-Z_{2}$ & $-Z_{3}$ & $0$ & $0$ & $-Z_{6}$ & $%
-Z_{7} $ & $0$ & $0$ \\ 
$\mathbf{Z}_{5}$ & $0$ & $Z_{3}$ & $-Z_{2}$ & $0$ & $0$ & $\left( 1-2\Omega
\right) Z_{7}$ & $\left( 1-2\Omega \right) Z_{6}$ & $-2\Omega Z_{8}$ & $%
2\Omega Z_{8}$ \\ 
$\mathbf{Z}_{6}$ & $2\Omega Z_{7}$ & $0$ & $0$ & $Z_{6}$ & $-\left(
1-2\Omega \right) Z_{7}$ & $0$ & $0$ & $2\Omega Z_{3}$ & $2\Omega Z_{2}$ \\ 
$\mathbf{Z}_{7}$ & $\,-2\Omega Z_{8}$ & $0$ & $0$ & $Z_{7}$ & $\,-\left(
1-2\Omega \right) Z_{6}$ & $0$ & $0$ & $-2\Omega Z_{2}$ & $-2\Omega Z_{3}$
\\ 
$\mathbf{Z}_{8}$ & $-2\Omega Z_{9}$ & $2\Omega Z_{7}$ & $2\Omega Z_{6}$ & $0$
& $2\Omega Z_{8}$ & $-2\Omega Z_{3}$ & $2\Omega Z_{2}$ & $0$ & $2\Omega
\left( Z_{1}+2\Omega Z_{5}\right) $ \\ 
$\mathbf{Z}_{9}$ & $2\Omega Z_{8}$ & $2\Omega Z_{6}$ & $2\Omega Z_{7}$ & $0$
& $-2\Omega Z_{8}$ & $-2\Omega Z_{2}$ & $2\Omega Z_{3}$ & $-2\Omega \left(
Z_{1}+2\Omega Z_{5}\right) $ & $0$ \\ \hline\hline
\end{tabular}%
\label{tab5}%
\end{table}
\end{landscape}%

\begin{landscape}
\begin{table}[tbp] \centering%
\caption{Adjoint representation for the Lie point symmetries of the
Shallow-water system for $\phi_{0}=\pi/2$, $sh$ indicates $sinh$ while $ch$
means $cosh$ (1/2) }%
\begin{tabular}{cccccc}
\hline\hline
$Ad\left( e^{\left( \varepsilon \mathbf{Z}_{i}\right) }\right) \mathbf{Z}%
_{j} $ & $\mathbf{Z}_{1}$ & $\mathbf{Z}_{2}$ & $\mathbf{Z}_{3}$ & $\mathbf{Z}%
_{4}$ & $\mathbf{Z}_{5}$ \\ \hline
$\mathbf{Z}_{1}$ & ${\small Z}_{1}$ & ${\small Z}_{2}$ & ${\small Z}_{3}$ & $%
{\small Z}_{4}$ & ${\small Z}_{5}$ \\ 
$\mathbf{Z}_{2}$ & ${\small Z}_{1}$ & ${\small Z}_{2}$ & ${\small Z}_{3}$ & $%
{\small Z}_{4}{\small -\varepsilon Z}_{2}$ & ${\small Z}_{5}{\small %
-\varepsilon Z}_{3}$ \\ 
$\mathbf{Z}_{3}$ & ${\small Z}_{1}$ & ${\small Z}_{2}$ & ${\small Z}_{3}$ & $%
{\small Z}_{4}{\small -\varepsilon Z}_{3}$ & ${\small Z}_{5}{\small %
+\varepsilon Z}_{2}$ \\ 
$\mathbf{Z}_{4}$ & ${\small Z}_{1}$ & ${\small e}^{\varepsilon }{\small Z}%
_{2}$ & ${\small e}^{\varepsilon }{\small Z}_{3}$ & ${\small Z}_{4}$ & $%
{\small Z}_{5}$ \\ 
$\mathbf{Z}_{5}$ & ${\small Z}_{1}$ & $\cos \left( \varepsilon \right) 
{\small Z}_{2}{\small +}\sin \left( \varepsilon \right) {\small Z}_{3}$ & $%
{\small -}\sin \left( \varepsilon \right) {\small Z}_{2}{\small +}\cos
\left( \varepsilon \right) {\small Z}_{3}$ & ${\small Z}_{4}$ & ${\small Z}%
_{5}$ \\ 
$\mathbf{Z}_{6}$ & ${\small Z}_{1}{\small -2\Omega \varepsilon Z}_{7}$ & $%
{\small Z}_{2}$ & ${\small Z}_{3}$ & ${\small Z}_{4}{\small -\varepsilon Z}%
_{6}$ & ${\small Z}_{5}$ \\ 
$\mathbf{Z}_{7}$ & ${\small Z}_{1}{\small +2\Omega \varepsilon Z}_{6}$ & $%
{\small Z}_{2}$ & ${\small Z}_{3}$ & ${\small Z}_{4}{\small -\varepsilon Z}%
_{7}$ & ${\small Z}_{5}$ \\ 
$\mathbf{Z}_{8}$ & ${\small ch}^{2}\left( 2\Omega \varepsilon \right) 
{\small Z}_{1}{\small -sh}^{2}\left( 2\Omega \varepsilon \right) {\small Z}%
_{5}{\small +}\frac{1}{2}{\small sh}\left( 4\Omega \varepsilon \right) 
{\small Z}_{9}$ & ${\small ch}\left( 2\Omega \varepsilon \right) {\small Z}%
_{2}{\small +sh}\left( 2\Omega \varepsilon \right) {\small Z}_{7}$ & $%
{\small ch}\left( 2\Omega \varepsilon \right) {\small Z}_{3}{\small +sh}%
\left( 2\Omega \varepsilon \right) {\small Z}_{6}$ & ${\small Z}_{4}$ & $%
{\small -}\frac{sh^{2}\left( 2\Omega \varepsilon \right) }{2\Omega }{\small Z%
}_{1}{\small +ch}^{2}\left( 2\Omega \varepsilon \right) {\small Z}_{5}%
{\small -}\frac{sh\left( 4\Omega \varepsilon \right) }{4\Omega }{\small Z}%
_{9}$ \\ 
$\mathbf{Z}_{9}$ & ${\small ch}^{22}\left( 2\Omega \varepsilon \right) 
{\small Z}_{1}{\small -sh}^{2}\left( 2\Omega \varepsilon \right) {\small Z}%
_{5}{\small +}\frac{1}{2}{\small sh}\left( 4\Omega \varepsilon \right) 
{\small Z}_{8}$ & ${\small ch}\left( 2\Omega \varepsilon \right) {\small Z}%
_{2}{\small +sh}\left( 2\Omega \varepsilon \right) {\small Z}_{6}$ & $%
{\small ch}\left( 2\Omega \varepsilon \right) {\small Z}_{3}{\small +sh}%
\left( 2\Omega \varepsilon \right) {\small Z}_{7}$ & ${\small Z}_{4}$ & $%
{\small -}\frac{sh^{2}\left( 2\Omega \varepsilon \right) }{2\Omega }{\small Z%
}_{1}{\small +ch}^{2}\left( 2\Omega \varepsilon \right) {\small Z}_{5}%
{\small +}\frac{sh\left( 4\Omega \varepsilon \right) }{4\Omega }{\small Z}%
_{8}$ \\ \hline\hline
\end{tabular}%
\label{tab6}%
\end{table}
\end{landscape}%

\begin{landscape}
\begin{table}[tbp] \centering%
\caption{Adjoint representation for the Lie point symmetries of the
Shallow-water system for $\phi_{0}=\pi/2$, $sh$ indicates $sinh$ while $ch$
means $cosh$ (2/2) }%
\begin{tabular}{ccccc}
\hline\hline
$Ad\left( e^{\left( \varepsilon \mathbf{Z}_{i}\right) }\right) \mathbf{Z}%
_{j} $ & $\mathbf{Z}_{6}$ & $\mathbf{Z}_{7}$ & $\mathbf{Z}_{8}$ & $\mathbf{Z}%
_{9}$ \\ \hline
$\mathbf{Z}_{1}$ & $\cos \left( 2\Omega \varepsilon \right) {\small Z}_{6}%
{\small +}\sin \left( 2\Omega \varepsilon \right) {\small Z}_{7}$ & ${\small %
-}\sin \left( 2\Omega \varepsilon \right) {\small Z}_{6}{\small +}\cos
\left( 2\Omega \varepsilon \right) {\small Z}_{7}$ & $\cos \left( 2\Omega
\varepsilon \right) {\small Z}_{8}{\small -}\sin \left( 2\Omega \varepsilon
\right) {\small Z}_{9}$ & $\sin \left( 2\Omega \varepsilon \right) {\small Z}%
_{8}{\small +}\cos \left( 2\Omega \varepsilon \right) {\small Z}_{9}$ \\ 
$\mathbf{Z}_{2}$ & ${\small Z}_{6}$ & ${\small Z}_{7}$ & ${\small Z}_{8}%
{\small +2\Omega \varepsilon Z}_{7}$ & ${\small Z}_{9}{\small +2\Omega
\varepsilon Z}_{6}$ \\ 
$\mathbf{Z}_{3}$ & ${\small Z}_{6}$ & ${\small Z}_{7}$ & ${\small Z}_{8}%
{\small -2\Omega \varepsilon Z}_{6}$ & ${\small Z}_{9}{\small +2\Omega
\varepsilon Z}_{7}$ \\ 
$\mathbf{Z}_{4}$ & ${\small e}^{\varepsilon }{\small Z}_{6}$ & ${\small e}%
^{\varepsilon }{\small Z}_{7}$ & ${\small Z}_{8}$ & ${\small Z}_{9}$ \\ 
$\mathbf{Z}_{5}$ & ${\small Z}_{6}$ & ${\small Z}_{7}$ & $\cos \left(
\varepsilon \right) {\small Z}_{8}{\small +}\sin \left( \varepsilon \right) 
{\small Z}_{9}$ & ${\small -}\sin \left( \varepsilon \right) {\small Z}_{8}%
{\small +}\cos \left( \varepsilon \right) {\small Z}_{9}$ \\ 
$\mathbf{Z}_{6}$ & ${\small Z}_{6}$ & ${\small Z}_{7}$ & ${\small Z}_{8}%
{\small -2\Omega \varepsilon Z}_{3}$ & ${\small Z}_{9}{\small +2\Omega
\varepsilon Z}_{2}$ \\ 
$\mathbf{Z}_{7}$ & ${\small Z}_{6}$ & ${\small Z}_{7}$ & ${\small Z}_{8}%
{\small +2\Omega \varepsilon Z}_{2}$ & ${\small Z}_{9}{\small +2\Omega
\varepsilon Z}_{3}$ \\ 
$\mathbf{Z}_{8}$ & ${\small sh}\left( 2\Omega \varepsilon \right) {\small Z}%
_{3}{\small +ch}\left( 2\Omega \varepsilon \right) {\small Z}_{6}$ & $%
{\small ch}\left( 2\Omega \varepsilon \right) {\small Z}_{7}{\small -sh}%
\left( 2\Omega \varepsilon \right) {\small Z}_{2}$ & ${\small Z}_{8}$ & $%
{\small ch}\left( 4\Omega \varepsilon \right) {\small Z}_{9}{\small +sh}%
\left( 4\Omega \varepsilon \right) \left( Z_{1}-2\Omega Z_{5}\right) $ \\ 
$\mathbf{Z}_{9}$ & ${\small -sh}\left( 2\Omega \varepsilon \right) {\small Z}%
_{2}{\small +ch}\left( 2\Omega \varepsilon \right) {\small Z}_{6}$ & $%
{\small ch}\left( 2\Omega \varepsilon \right) {\small Z}_{7}{\small -sh}%
\left( 2\Omega \varepsilon \right) {\small Z}_{3}$ & ${\small ch}\left(
4\Omega \varepsilon \right) {\small Z}_{8}{\small +sh}\left( 4\Omega
\varepsilon \right) \left( Z_{1}-2\Omega Z_{5}\right) $ & ${\small Z}_{9}$
\\ \hline\hline
\end{tabular}%
\label{tab6b}%
\end{table}
\end{landscape}%

\section{Similarity solutions}

\label{sec4}

In this Section we apply the Lie point symmetries to determine similarity
solutions for the Shallow-water system, (\ref{sw.01}), (\ref{sw.02}) and (%
\ref{sw.03}). For arbitrary latitude the admitted Lie point symmetries
provide only travelling-wave solutions. Therefore we focus on the case with $%
\phi _{0}=0$, where there are additional Lie point symmetries which can
provide alternate reductions.

In order to determine exact similarity solutions for the original system we
need the application of at least two Lie point symmetries. Thus in the
following lines we continue with the investigation of exact similarity
solutions by using the Lie invariants of the following set of Lie point
symmetries $\left\{ X_{1}+X_{2}+X_{3},X_{2}-X_{3}\right\} ,$ and in the
limit at the equator we study similarity solutions from the Lie symmetries $%
\left\{ Y_{2},Y_{5}\right\} ,~\left\{ Y_{2},Y_{5}\right\} ,~\left\{
Y_{4},Y_{5}\right\} $. The reduced equations are ordinary differential
equations and lead to closed-form solutions.

\subsection{Reduction with $\left\{ X_{1}+X_{2}+X_{3},X_{2}-X_{3}\right\} $}

Consider the similarity transformation provided by the set of the symmetry
vectors $\left\{ X_{1}+X_{2}+X_{3},X_{2}-X_{3}\right\} $, 
\begin{equation}
h\left( t,x,y\right) =H\left( w\right) ~,~u\left( t,x,y\right) =U\left(
w\right) ,~v\left( t,x,y\right) =V\left( w\right) \text{ with }w=\left(
x+y\right) -2t,  \label{st.01}
\end{equation}%
while the reduced system is%
\begin{equation}
\frac{G\left( w\right) }{2\Omega _{z}}\frac{{H}_{w}}{H}={V^{2}-2V-U^{2}+2%
\,U+\Omega _{y}H}^{2}{\,U+\Omega _{y}\,H}^{2}{V,}  \label{st.02}
\end{equation}

\begin{equation}
-\frac{G\left( w\right) }{2\Omega _{z}}{U}_{w}=\,2HU\Omega _{y}\left(
HV+HU+1\right) -VHg-UHg+U^{2}V+2\,UV^{2}-4\,UV+4\,V+V^{3}-4V^{2},
\label{st.03}
\end{equation}

\begin{eqnarray}
-\frac{G\left( w\right) }{2\Omega _{z}}{V}_{w}
&=&VHg+UHg-2U^{2}V-UV^{2}+4\,UV-4\,U+4U^{2}-U^{3}+  \label{st.04} \\
&&-H\Omega _{y}\left( U^{2}H+2U\left( 1+H\right) +VH\left( 2-V\right)
\right) ,  \notag
\end{eqnarray}%
where%
\begin{eqnarray}
G\left( w\right) &=&6U^{2}+6V^{2}-12\,V-4\,Hg+12UV+8-3U^{2}V-3UV^{2}  \notag
\\
&&+2UHg-V^{3}-U^{3}+2\,VHg-12\,U+  \notag \\
&&-H\Omega _{y}\left( U+V-2\right) \left( U+3HU+H\left( 2-V-H\Omega
_{z}\right) +2-V\right) .
\end{eqnarray}

\begin{figure}[tbp]
\centering\includegraphics[scale=0.5]{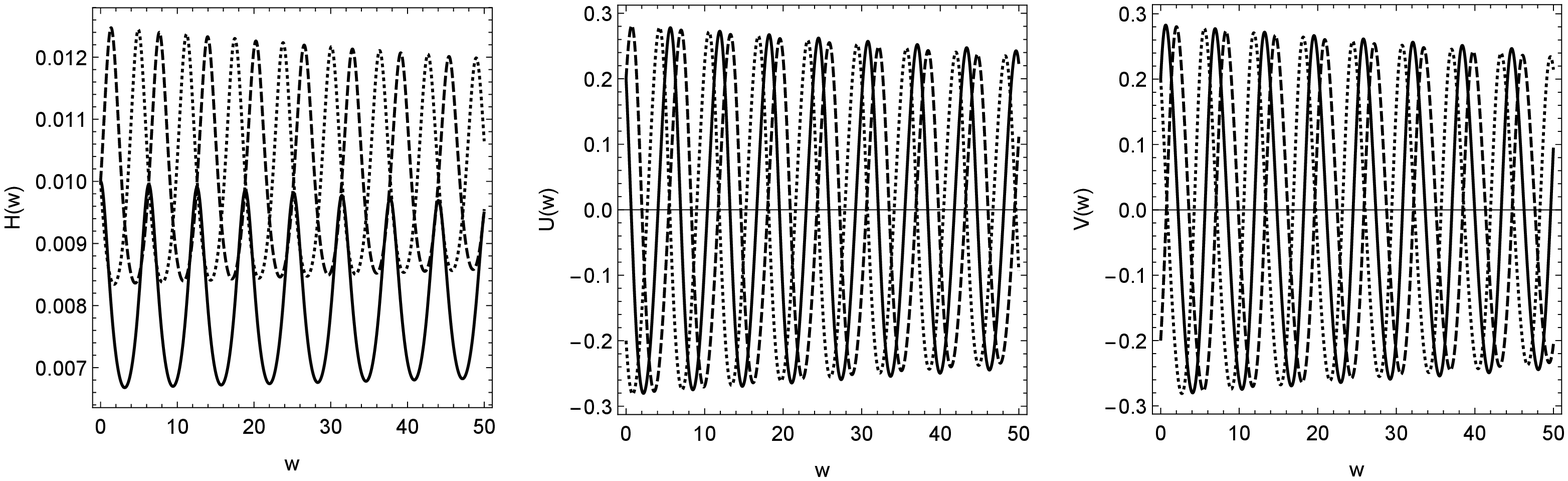}
\caption{Qualitative evolution of $H\left( w\right) ,~U\left( w\right)
,~V\left( w\right) $ as it is given from numerical simulation of the
dynamical system (\protect\ref{st.02}), (\protect\ref{st.03}) and (\protect
\ref{st.03}). The plots are for $\Omega _{z}=\Omega _{y}=1$ and $g=10$ and
for different initial conditions as they are presented at the plots. }
\label{plot05}
\end{figure}

In Fig. \ref{plot05} we present the numerical solution of the
travelling-wave solution as it is given by the dynamical system (\ref{st.02}%
), (\ref{st.03}) and (\ref{st.03}). The numerical solution is for $\phi _{0}=%
\frac{\pi }{4}$ and $\Omega _{z}=\Omega _{y}$.

\subsection{Reduction with $\left\{ Y_{2},Y_{5}\right\} $}

The application of the Lie point symmetries $\left\{ Y_{2},Y_{5}\right\} $
lead to the invariants 
\begin{equation}
h\left( t,x,y\right) =H\left( t\right) ~,~u\left( t,x,y\right) =U\left(
t\right) ~,~v\left( t,x,y\right) =\frac{y}{t}+V\left( t\right) ,
\end{equation}%
where the unknown functions, $H\left( t\right) ,~U\left( t\right)
\,,~V\left( t\right) $, satisfy the following system%
\begin{equation}
tH_{t}+H=0~,~tU_{t}-\Omega H=0~,~tV_{t}+V=0,
\end{equation}%
with exact solution%
\begin{equation}
H\left( t\right) =\frac{H_{0}}{t}~,~U\left( t\right) =-\frac{H_{0}\Omega }{t}%
+U_{0}~,~V\left( t\right) =\frac{V_{0}}{t}.
\end{equation}

\subsection{Reduction with $\left\{ Y_{4},Y_{5}\right\} $}

Reduction with the Lie point symmetries $\left\{ Y_{4},Y_{5}\right\} $
provides the following system of ordinary differential equations%
\begin{eqnarray}
\frac{H_{w}}{H} &=&\frac{\Omega H^{2}+w-U}{L\left( w\right) },~  \label{e1}
\\
U_{w} &=&\frac{H\left( 2UH\Omega +w\Omega -g\right) }{L\left( w\right) }~,~
\label{e2} \\
V_{w} &=&V\left( U-w\right) ^{-1},  \label{e3}
\end{eqnarray}%
where 
\begin{equation}
h\left( t,x,y\right) =H\left( w\right) ~,~u\left( t,x,y\right) =U\left(
w\right) \text{ , }v\left( t,x,y\right) =\frac{y}{t}+V\left( w\right) ~\text{%
and}~w=\frac{x}{t}
\end{equation}%
and%
\begin{equation}
L\left( w\right) =wH\Omega \left( H+1\right) +\Omega UH^{2}-gH+\left(
w-U\right) ^{2}.
\end{equation}

However, that solution is not accepted because it leads to an infinite value
of $V\left( w\right) $. The qualitative evolution of $U\left( w\right) $ and 
$V\left( w\right) $ for various initial conditions is presented in Fig. \ref%
{plot1}

\begin{figure}[tbp]
\centering\includegraphics[scale=0.5]{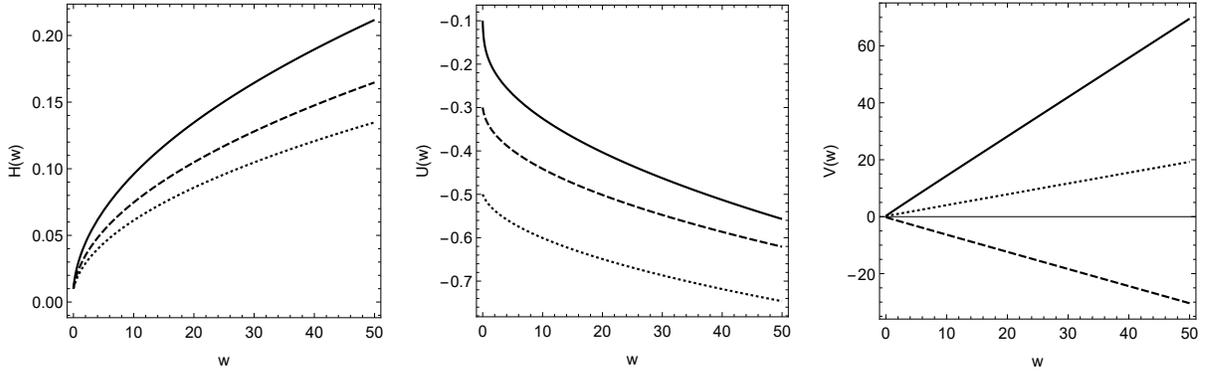}
\caption{Qualitative evolution of $H\left( w\right) ,~U\left( w\right) $ and 
$V\left( w\right) $ as they are given from the differential equation (%
\protect\ref{e5}) for $\Omega =1$ and for different values for the initial
conditions are they are presented in figures.}
\label{plot1}
\end{figure}

\section{Conclusions}

\label{sec5}

In this work we studied the group properties of the shallow-water equations
on a linear plane with the complete Coriolis force. In particular we
performed a classification of the admitted Lie point symmetries which are
the generators of one-parameter point transformations which leave invariant
shallow-water equations.

The coordinate system has been selected such that two terms of the Coriolis
force are involved in the shallow-water equations. These terms which are the
Coriolis components on the axes depend upon the position on the plane. For
an arbitrary position and for the more general form the shallow-water
equations admit three Lie point symmetries which are the time translation
and the two space translations. For specific values of the Coriolis terms
new symmetries follow.

When $\phi _{0}=0$, that is, at the equator, we found that the resulting
system of hyperbolic differential equations is invariant under the action of
a five-dimensional Lie algebra, while, when $\phi _{0}=\frac{\pi }{2}$, that
is, at the pole, the shallow-water equations admit nine Lie point
symmetries. In the latter scenario the Lie point symmetries form the same
algebra as the shallow-water equations without the Coriolis term. However,
the two algebras are in different representations. This case has been
studied before in the literature so we focused upon the case with $\phi
_{0}=0$.

For the latter case we applied the new Lie point symmetries to determine
similarity transformations and to write similarity solutions for the
equations by using closed-form functions. We presented reductions which
describe travelling-wave and scaling solutions. When it is possible, we
write the closed-form solution of the shallow-water equations. However, in
the remaining cases we solve numerically the reduced system and we study the
qualitative behaviour of the physical variables.

This work contributes to the subject of the group properties of hyperbolic
equations with emphasis in fluid dynamics. We present for the first time new
solutions of the shallow-wave equations with a complete Coriolis term. Such
an analysis shows the novelty and the power of the Lie symmetries for the
study of nonlinear phenomena.

In a future work we plan to extend our analysis in the case of nonlinear
topographies and with possible sources.

\end{document}